\documentclass[aps,showpacs,pra,tighten,amsmath,amssymb,groupaddress,twocolumn]{revtex4-1}
\usepackage{graphicx}
\usepackage{braket}
\usepackage{amsmath}
\usepackage{bm}
\usepackage{subfigure}
\begin{document}
\title{Optimal control for non-Markovian open quantum systems}
\author{Bin Hwang}
\author{Hsi-Sheng Goan}\email[Corresponding author: ]{goan@phys.ntu.edu.tw}
\affiliation{Department of Physics and Center for Theoretical Sciences, National Taiwan University, Taipei 10617, Taiwan}
\affiliation{Center for Quantum Science and Engineering, National Taiwan University, Taipei 10617, Taiwan}
\date{\today}
\begin{abstract}
An efficient optimal control theory based on the Krotov 
method is introduced for 
a non-Markovian open
quantum system with a time-nonlocal master equation in which the control
parameter and the bath correlation function are correlated. This
optimal control
method is developed via a quantum dissipation formulation
that transforms the time-nonlocal master equation to a set of
coupled linear time-local equations of motion 
in an extended auxiliary Liouville space. 
As an illustration, the optimal
control method is applied to find the control sequences for 
high-fidelity $Z$-gates and identity-gates of a qubit embedded
in a non-Markovian bath.  
$Z$-gates and identity-gates with errors less than
$10^{-5}$ for a wide range of bath decoherence
parameters can be achieved 
for the
non-Markovian open qubit system with control over only the $\sigma_z$
term. The control-dissipation correlation,
and the memory effect of the bath are crucial in achieving the
high-fidelity gates. 
\end{abstract}


\maketitle
\section{Introduction}

Quantum optimal control theory (QOCT) \cite{Rabitz88,Tannor92,Kosloff02,Montangero2007,Nielsen08,Khaneja05,Tannor11}
is a powerful tool that 
provides a variational
framework for calculating the optimal shaped pulse to
maximize a desired physical objective (or minimize a physical cost function).
It has been applied to various open quantum systems or models to
obtain control sequences for quantum gate operations 
\cite{Rebentrost2009,potz2008,potz2009,Jirari2009,Gordon08,Clausen10}.
Compared to the dynamical-decoupling-based method \cite{West10} 
in which a succession of
short and strong 
pulses designed to suppress decoherence is applied to the system, 
QOCT is a continuous dynamical modulation with many degrees of freedom for
selecting arbitrary shapes, durations and strengths for 
time-dependent control, 
and thus allows
significant reduction of the applied control energy and the
corresponding quantum gate error.  
The authors of Ref.~\cite{Rebentrost2009} investigated the
optimal control of a qubit coupled
to a two-level system that is exposed to a Markovian heat bath.
Although this may mimic the reduced non-Markovian dynamics of the qubit, it
is by no means a model of a qubit coupled directly to a non-Markovian
environment.  
The authors of Refs.~\cite{potz2008,potz2009,Jirari2009} 
investigated optimal quantum gate operations in the presence of
non-Markovian environments.  
However, to combine QOCT with a
non-Markovian master equation involving
time-ordered integration of the nonunitary (dissipation) terms
for noncommuting system and control operators, 
and for a nonlocal-in-time memory kernel, 
the numerical treatment is rather
mathematically involved and 
computationally demanding. 
All of the QOCT approaches mentioned above
\cite{Rebentrost2009,potz2008,potz2009,Jirari2009,Gordon08,Clausen10}
for open quantum systems employed
gradient-based \cite{Rabitz88,Khaneja05} 
algorithms for optimization.

A somewhat different QOCT approach from the standard gradient
optimization methods 
is the Krotov iterative method 
\cite{krotov1996,Tannor92,Nielsen08,Tannor11}. 
The Krotov method has several
appealing advantages \cite{Tannor92,Nielsen08,Tannor11}
over the gradient methods: 
(a) monotonic increase of the objective 
with iteration number,
(b) no requirement for a line search,  and (c) macrosteps
at each iteration.
A version of the Krotov
optimization method has been used recently  in 
Ref.~\cite{Schmidt11} to deal with the non-Markovian
optimal control problem of a quantum Brownian motion model with an exact
stochastic equation of motion (master equation).
We note however that only a few
non-Markovian open quantum system models are exactly solvable
\cite{Schmidt11,Breuer02,Hu92,Diosi98,Strunz04,Yan09,Chang10,Yu10}
(the quantum Brownian model is one of them),
and the exact master equations of these exactly solvable 
models are known to be in a time-local
(time-convolutionless) form with time-dependent decoherence or decay
rates, without involving the time-ordering problem of non-commuting
operators. \cite{Schmidt11,Breuer02,Hu92,Diosi98,Strunz04,Yan09,Chang10,Yu10}.  
With time-local equations of motion, the Krotov method could be
directly employed to deal with optimal-control problems. 
Although it is commendable to derive an exact master equation, not too many
problems can be exactly worked out in this way. 
For example, no exact
master equation can be derived for the non-Markovian
qubit-environment (spin-boson) model studied as an illustration for quantum gate
operations in this paper. It is thus
important that an efficient QOCT approach based on the Krotov
method for perturbative treatment of general (not limited to
just some certain classes of) 
non-Markovian open quantum systems should be developed. 
The
perturbative non-Markovian master equation under only the Born
approximation (or the weak system-bath coupling approximation)
\cite{Breuer02} is in the form of a 
time-ordered non-commuting integro-differential equation [e.g., see Eqs.~(\ref{EOM0}) and (\ref{K_diss})], and  at first sight it is
not at all clear
how to effectively combine this kind of master equation with the
Krotov optimization method. 
The study
presented in this paper provides just such an efficient 
QOCT approach based on the Krotov method to
deal with time-nonlocal non-Markovian open quantum systems. 
Our approach 
transforms the
 time-ordered non-commuting integro-differential master equation into
 a set of time-local coupled differential equations with 
the small price of introducing auxiliary density matrices 
in an extended auxiliary Liouville space. 
As a result, 
incorporation of the resultant time-local equations with the Krotov
optimization method becomes effective. 
We then apply the developed Krotov method 
to the problem of finding the optimal quantum gate control sequence for 
a qubit in a non-Markovian environment (bath).
Our results illustrate 
that the control parameter can be engineered
to efficiently counteract and suppress the environment effect for 
non-Markovian open systems 
with long bath correlation times (long memory effects).
We also find that high-fidelity quantum gates with error smaller than 
$10^{-5}$ can be achieved at long gate operation times for a wide
range of bath decoherence parameters. 
This is in contrast to the cases in the literature
\cite{Rebentrost2009,potz2008,potz2009,Jirari2009}, 
where the non-Markovian
systems were mainly studied in a parameter regime very close to Markovian
systems, and thus
no significant reduction of the quantum gate errors was observed.

\section{Non-Markovian master equation and quantum optimal control theory}
In realistic experiments, one may have only limited control over
the system Hamiltonian, and the maximum control parameter strength that can be
realized 
is also restricted. 
Let us consider a total system with Hamiltonian   
$H=H_S(t)+H_I+H_B$. Here the system Hamiltonian
$H_S(t)=-\hbar\varepsilon(t)\sigma_z/2-\hbar\Omega\sigma_x/2$,
describes a qubit with 
a time-dependent control parameter $\varepsilon(t)$ and 
a fixed tunneling frequency $\Omega$, 
i.e., having control over only the $\sigma_z$ term. 
The bath Hamiltonian is 
$H_B=\sum_q\hbar\omega_q{b_q}^\dagger b_q$,
where ${b_q}^\dagger$ (${b_q}$) is the creation (annihilation) operator
of the bath mode $q$ with frequency $\omega_q$. The interaction
Hamiltonian $H_I$ between the system and the bath without
making the rotating-wave approximation is of the form
$H_I=\sigma_x\sum_q c_q(b_q+{b_q}^\dagger)$,
where $c_q$ is the coupling constant of bath mode $q$ to the qubit system. 
Following the standard perturbation theory,
we obtain (see the Appendix \ref{appendix} for the derivation)
under only the Born approximation the 
time-convolution master equation for the reduced system density matrix as 
\cite{Meier99,Yan01,Kleinekathofer04,Kleinekathofer06}
\begin{equation}
\dot\rho(t)=\mathcal{L}_S(t)\rho(t)+
\{\mathcal{L}_x\mathcal{K}(t)+[\mathcal{L}_x\mathcal{K}(t)]^\dagger\}.
\label{EOM0} 
\end{equation}
Here the superoperators $\mathcal{L}_S(t)$ and $\mathcal{L}_x$ are
defined via their actions on an arbitrary operator $A$, respectively,
as $\mathcal{L}_S(t)A=\frac{1}{i\hbar}[H_S(t),A]$ and
$\mathcal{L}_xA=\frac{1}{i\hbar}[\sigma_x,A]$.  
The non-Hermitian (dissipation) operator can be written as 
\begin{equation}
\mathcal{K}(t)=\frac{1}{i\hbar}\int_0^t dt'C(t-t')\mathcal{U}_S(t,t')\sigma_x \rho(t'),
\label{K_diss}
\end{equation}
where the unitary qubit system propagator superoperator $\mathcal{U}_S(t,t')=T_+
\exp\{\int^t_{t'}d\tau\mathcal{L}_S(\tau)\}$ 
with $T_+$ being the time-ordering operator, 
and the bath correlation function (CF) is
$C(t-t')=\int_0^\infty d\omega J(\omega)\cos[\omega
(t-t')]\coth[{\hbar\omega}/({2k_BT})] -i\int_0^\infty d\omega
J(\omega)\sin[\omega (t-t')]$
with $T$ being the temperature. 
Note that Eq.~(\ref{K_diss}) contains the bath CF
$C(t,t')$ and the time-ordered system propagator
superoperator $\mathcal{U}_S(t,t')$ which involves the control parameter
$\varepsilon(t)$ through $H_S(t)$ in $\mathcal{L}_S(t)$. Thus the
control parameter and 
bath-induced nonunitary (dissipation)
effect are correlated. This paves the way for 
manipulating the control sequence to counteract the effect of the
bath on the system dynamics. 
This coherent control of non-Markovian open quantum systems
is in contrast to various Markovian
approaches of engineering reservoirs \cite{ER} and incoherent controls by
directly manipulating the environments \cite{ICE}.

In the framework of the QOCT, 
one would like to maximize the quality 
(fidelity) value of some target at
time $t_f$. 
Suppose the desired state-independent unitary
quantum gate operation is denoted as $\mathcal{O}$
and the target is to perform a state-independent
quantum gate operation as close as possible to $\mathcal{O}$.
We choose the trace distance
between the desired target superoperator $\mathcal{O}$
and the actual nonunitary propagator superoperator $\mathcal{X}(t_f)$
at the final operation time $t_f$
to characterize the gate error, i.e.,
$\text{(error)}=\text{Tr}\{[\mathcal{O}-\mathcal{X}(t_f)]^2\}/N$,
where $N$ is the dimension value of $\mathcal{O}$.
As minimizing the trace distance is similar to 
maximizing the real part of the 
trace fidelity \cite{Rebentrost2009},
we choose  
$\mathcal{F}={\rm Re}[\text{Tr}\{\mathcal{O}^\dagger\mathcal{X}(t_f)\}]/N$
as a quality (fidelity) measure of how well 
$\mathcal{X}(t_f)$ 
approaches the target $\mathcal{O}$ in the QOCT framework.
In realistic control problems, 
it is desirable that the optimal control sequence can provide highest
quality (fidelity) with minimum energy consumption. 
Thus we introduce an objective function of the form 
\begin{equation}
\mathcal{J}=\mathcal{F}-\int_0^{t_f} dt'\lambda(t')[\varepsilon(t')-\varepsilon_0(t')]^2,
\label{Costfx}
\end{equation}
where $\lambda(t)$ is a positive function that can be adjusted and chosen
empirically, $\varepsilon(t)$ is the control parameter and
$\varepsilon_0(t)$ is a reference control value that can be properly
chosen \cite{Tannor92,Kosloff02}.  
Then the task of quantum control is to maximize
the objective (\ref{Costfx}) under the constraint of the 
equation of motion of $\mathcal{X}(t)$ obtained by 
replacing $\rho(t)$ with $\mathcal{X}(t)$ in Eqs.~(\ref{EOM0})
and (\ref{K_diss}).

The time-dependent control parameter $\varepsilon(t)$ enters into
the exponent of the time-ordered system 
propagator $\mathcal{U}_S(t,t')$, 
and thus appears inside the memory kernel or dissipation operator 
Eq.~(\ref{K_diss}).  
As a result, Eq.~(\ref{EOM0}) is a nonlocal time-ordered
integro-differential equation in which values of the control parameter at
all earlier times come into play, and is difficult to incorporate 
within the framework of QOCT. 
Thus an approach that retains the 
merits of the Krotov optimization method 
and removes the time-ordering and nonlocal
problems in the equation of motion is very much desired.

\section{Master equation and optimal control in extended Liouville space}
One important observation to deal with the time-nonlocal 
non-Markovian quantum master equation 
is to express the bath CF's in a multi-exponential form  
\cite{Meier99,Yan01,Kleinekathofer04,Kleinekathofer06},
$C(t-t')=\sum_j C_j(0)e^{\gamma_j(t-t')}$,
where $C_j(0)$ and $\gamma_j$ are complex constants and can be found
by numerical methods. 
Then 
Eq.~(\ref{K_diss})
can be written
as $\mathcal{K}(t)=\sum_j\mathcal{K}_j(t)$, where
$\mathcal{K}_j(t)=\frac{1}{i\hbar}\int_0^t
dt'C_j(0)e^{\gamma_j(t-t')}\mathcal{U}_S(t-t')\sigma_x  \rho(t')$.
Although $\mathcal{K}_j(t)$ is still a time-nonlocal and time-ordered
integration for non-commuting operators, 
if one takes the time derivative of $\mathcal{K}_j(t)$,
one obtains
\begin{equation}
\dot{\mathcal{K}}_j(t)=({1}/{i\hbar})C_j(0)\sigma_x \rho(t)+[\mathcal{L}_S(t)+\gamma_j]\mathcal{K}_j(t),
\label{EOM1}
\end{equation}
with initial condition $\mathcal{K}_j(0)=0$.
The same process can be done for the Hermitian conjugate 
${\mathcal{K}}^\dagger(t)\equiv\sum_j {\mathcal{K}_j}^\dagger(t)$. 
Equation (\ref{EOM0}) combined with Eq.~(\ref{EOM1}) and  
its Hermitian conjugate 
form a set of coupled linear equations of
motion that can be written as 
$\dot{\vec{\rho}}(t)=\hat\Lambda(t)\vec{\rho}(t)$,
 in terms of 
$\vec{\rho}(t)\equiv\{\rho(t),\mathcal{K}_j,{\mathcal{K}_j}^\dagger; j=1,2,3,...\}$ 
in an extended auxiliary Liouville space  
\cite{Meier99,Yan01,Kleinekathofer04,Kleinekathofer06}.
Obviously, the above equations are time-local and have no
time-ordering and memory kernel integration problems. 
This yields a simple, fast and stable iterative scheme to 
incorporate with the Krotov method. 
The formal solution of $\vec{\rho}(t)$ 
can be written as  
$\vec{\rho}(t)=\hat{\mathcal{G}}(t)\vec{\rho}(0)$,
where the associated propagator superoperator can be shown to satisfy
$[{\partial\hat{\mathcal{G}}(t)}/{\partial t}]=\hat\Lambda(t)\hat{\mathcal{G}}(t)$
with $\hat{\mathcal{G}}(0)=\hat{{I}}_\mathcal{N}$.
Here 
$\hat{{I}}_\mathcal{N}$ is the identity operator in the extended
Liouville space and $\mathcal{N}$ is the dimension of
$\hat{\mathcal{G}}(t)$.  
The real part of the trace fidelity 
for the propagator 
$\hat{\mathcal{G}}(t_f)$ is
$\mathcal{F}={\rm Re}[\text{Tr}\{\hat{\mathcal{Q}}^\dagger\hat{\mathcal{G}}(t_f)\}]/\mathcal{N}$,
where $\hat{\mathcal{Q}}$ is the target operator $\mathcal{O}$ in the
extended Liouville space. 
The goal of quantum optimal control here is to reach a desired
target $\hat{\mathcal{Q}}$
with maximum objective function $\mathcal{J}$ (or fidelity
$\mathcal{F}$) in a certain time $t_f$. 
The optimal algorithm following the Krotov method \cite{krotov1996} is
summarized as follows \cite{Tannor92,Kosloff02,Nielsen08}.
({i}) Guess an initial control sequence $\varepsilon_0(t)$.
({ii}) Use the equations of motion to find the forward
propagator $\hat{\mathcal G}_k(t)$ with the initial condition
$\hat{\mathcal G}(0)=\hat{{I}}_{\mathcal{N}}$ ($k=0$ for the first iteration).
({iii}) Find an auxiliary backward propagator
$\hat{\mathcal B}_k(t)$ with the condition $\hat{\mathcal
  B}(t_f)=\hat{\mathcal {Q}}^\dagger$. 
({iv}) Propagate $\hat{\mathcal G}_{k+1}(t)$ again forward in time,
and update the control parameter 
iteratively with the rule
%
$\varepsilon_{k+1}(t)=
\varepsilon_{k}(t)+\frac{1}{2\lambda(t)}\text{Re}[\text{Tr}\{\hat{\mathcal
  B}_{k}(t)\frac{\partial \Lambda(t)}{\partial
  \varepsilon(t)}\hat{\mathcal G}_{k+1}(t)\}].
$
({v}) Repeat steps (iii) and (iv) until a preset fidelity (error)
is reached or until a given number of iterations has been performed.

\section{Results and discussions} 
To find the optimal control sequence for a state-independent single-qubit
gates in a non-Markovian bath, one needs to know the bath spectral density to
calculate the bath CF. We can deal with any form of the
bath spectral density, but for simplicity, we take an Ohmic spectral
density in the form of $J(\omega)=\alpha\omega e^{-\omega/\omega_c}$, where
$\alpha$ is a dimensionless damping constant, and $\omega_c$ is the
bath cutoff frequency. 
The bath CF 
can thus be calculated 
and can then be expanded directly 
in a multi-exponential form. 
The values of $C_j(0)$ and $\gamma_j$ of the exponentials  
are obtained numerically 
with the requirement that 
the error
between the actual CF 
and the
approximated CF  
is chosen to be less than or equal to $10^{-7}$.
Only three or four exponential terms in the expansion 
are required
to express the bath CF
with such high accuracy
as compared to a sum of more than 48 exponentials 
needed to express the same bath CF at a low temperature of $T=0.2\Omega$
through the spectral density parametrization \cite{Meier99}. 
We note here that
both the
parametrization of Ref.~\cite{Meier99}, which is highly efficient at
moderate to high temperatures, and the direct bath CF
decomposition of our approach are 
well-suited for the parametrization of highly structured spectral
densities, leading to long and oscillatory bath CF's.
However, our approach has a great computational advantage over the commonly
used approach of spectral density parametrization 
\cite{Meier99,Yan01,Kleinekathofer04,Kleinekathofer06}
at low temperatures.

\begin{figure}[tbp]
\mbox{\subfigure{\includegraphics[width=0.45\linewidth]{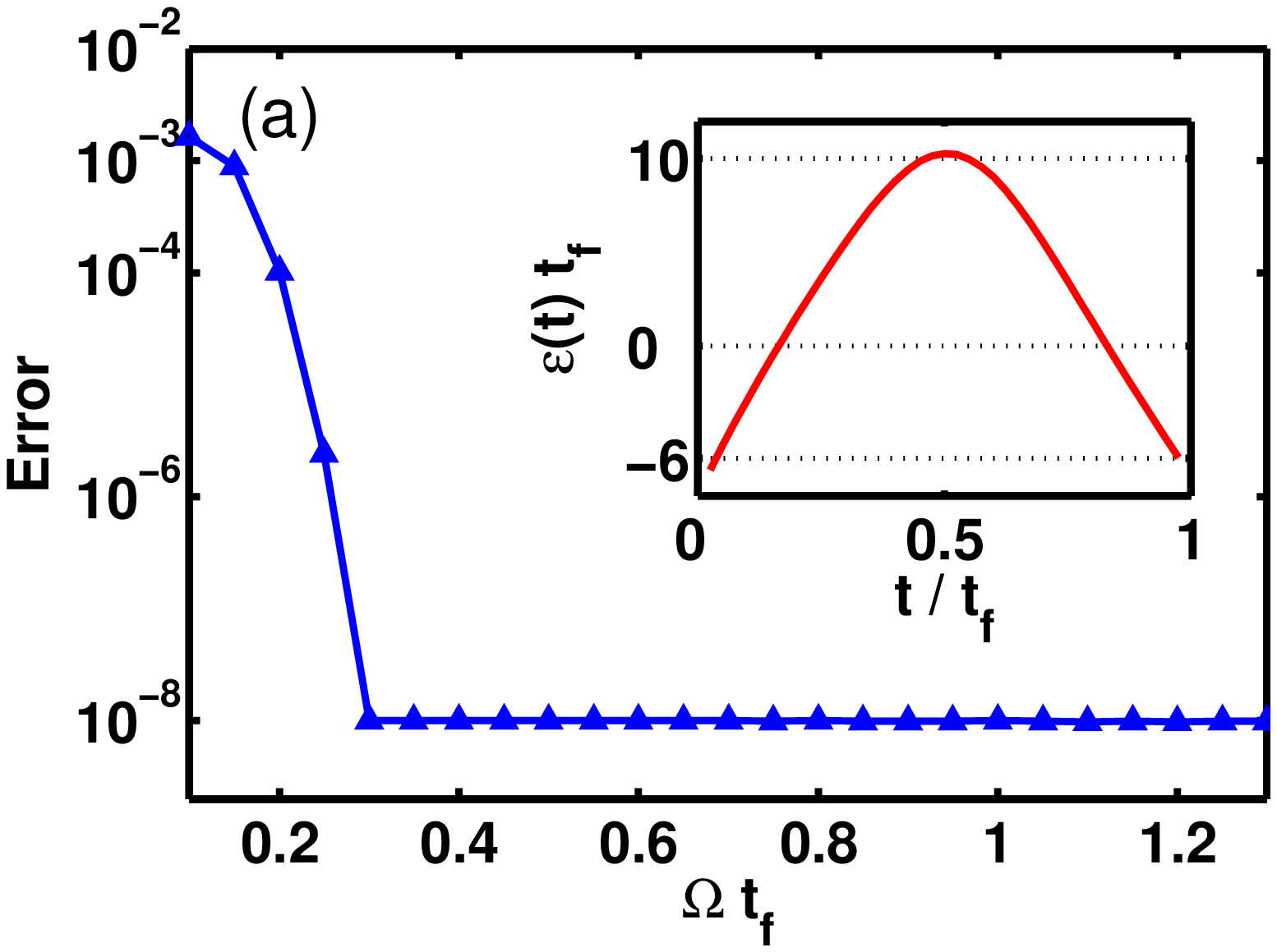}}\subfigure{\includegraphics[width=0.45\linewidth]{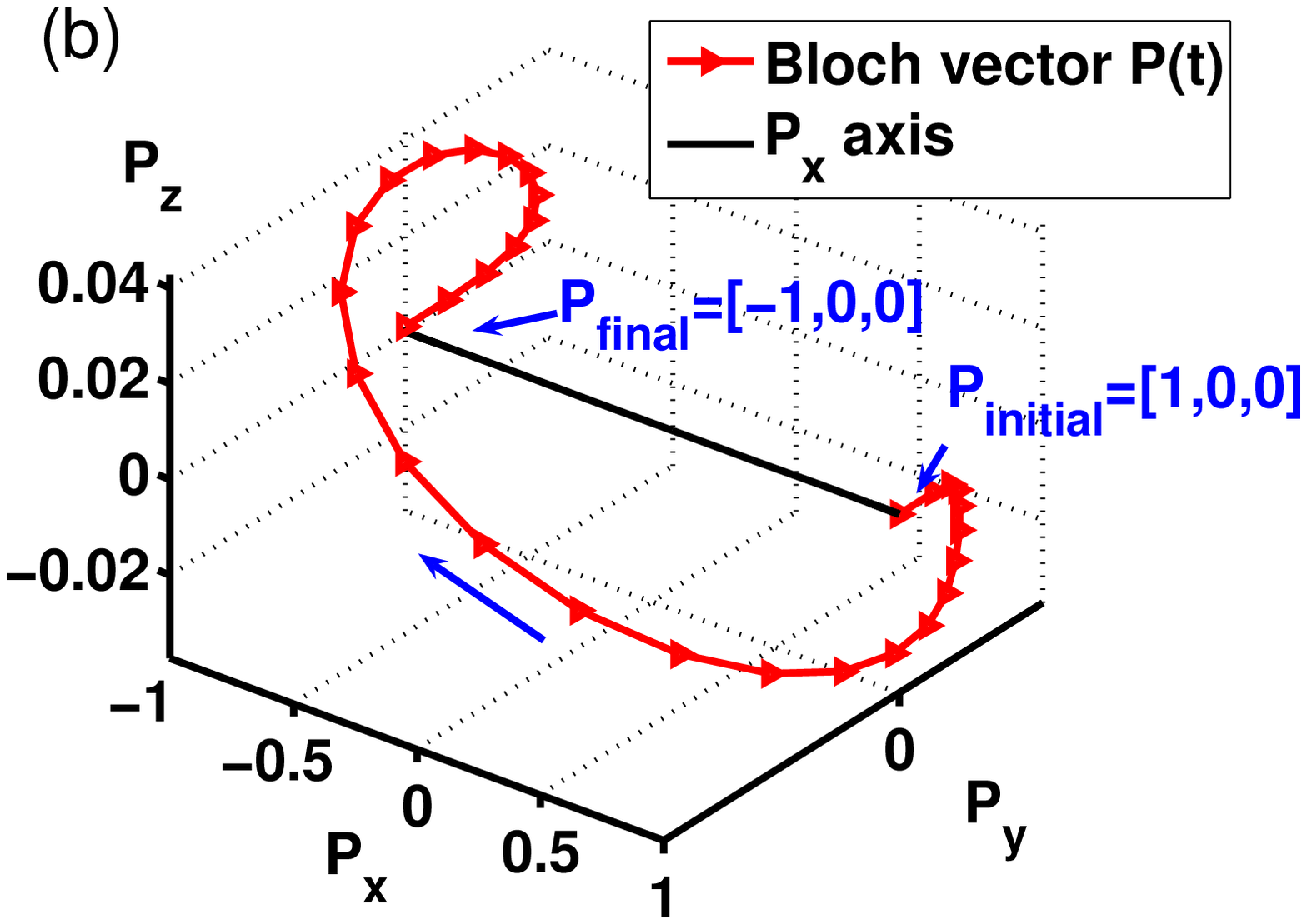}}}
\caption{(Color online)
(a) Unitary $Z$-gate error versus operation time $t_f$. (b) Bloch vector
trajectory of an optimal unitary 
$Z$-gate operation with the initial polarization condition $P(0)=[1,0,0]$.
The inset in (a) is the optimal control pulse sequence
for any $t_f\geq 0.3 \Omega$.}
\label{ideal gate}
\end{figure}

\begin{figure}[tbp]
\mbox{\subfigure{\includegraphics[width=0.45\linewidth]{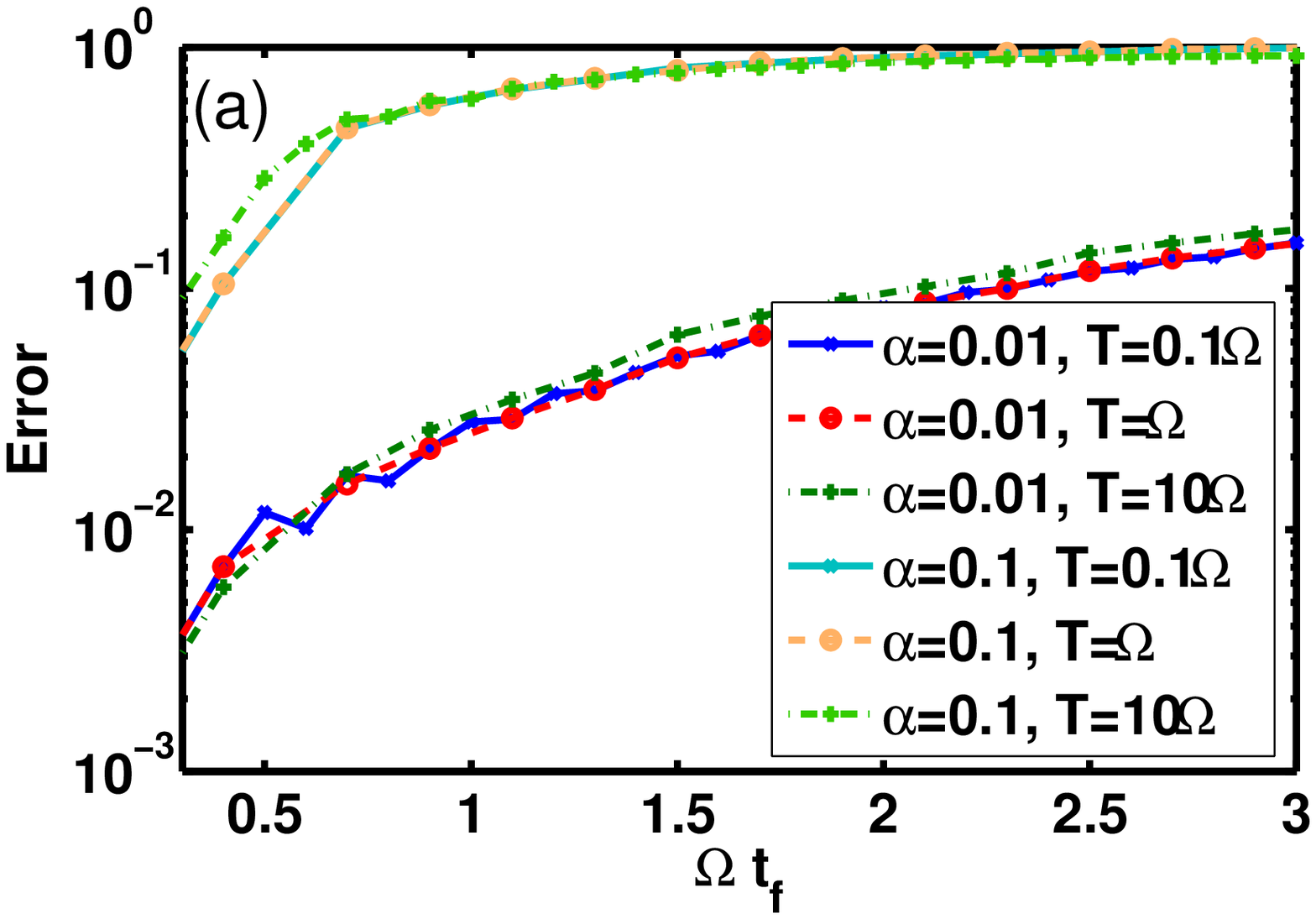}\subfigure{\includegraphics[width=0.45\linewidth]{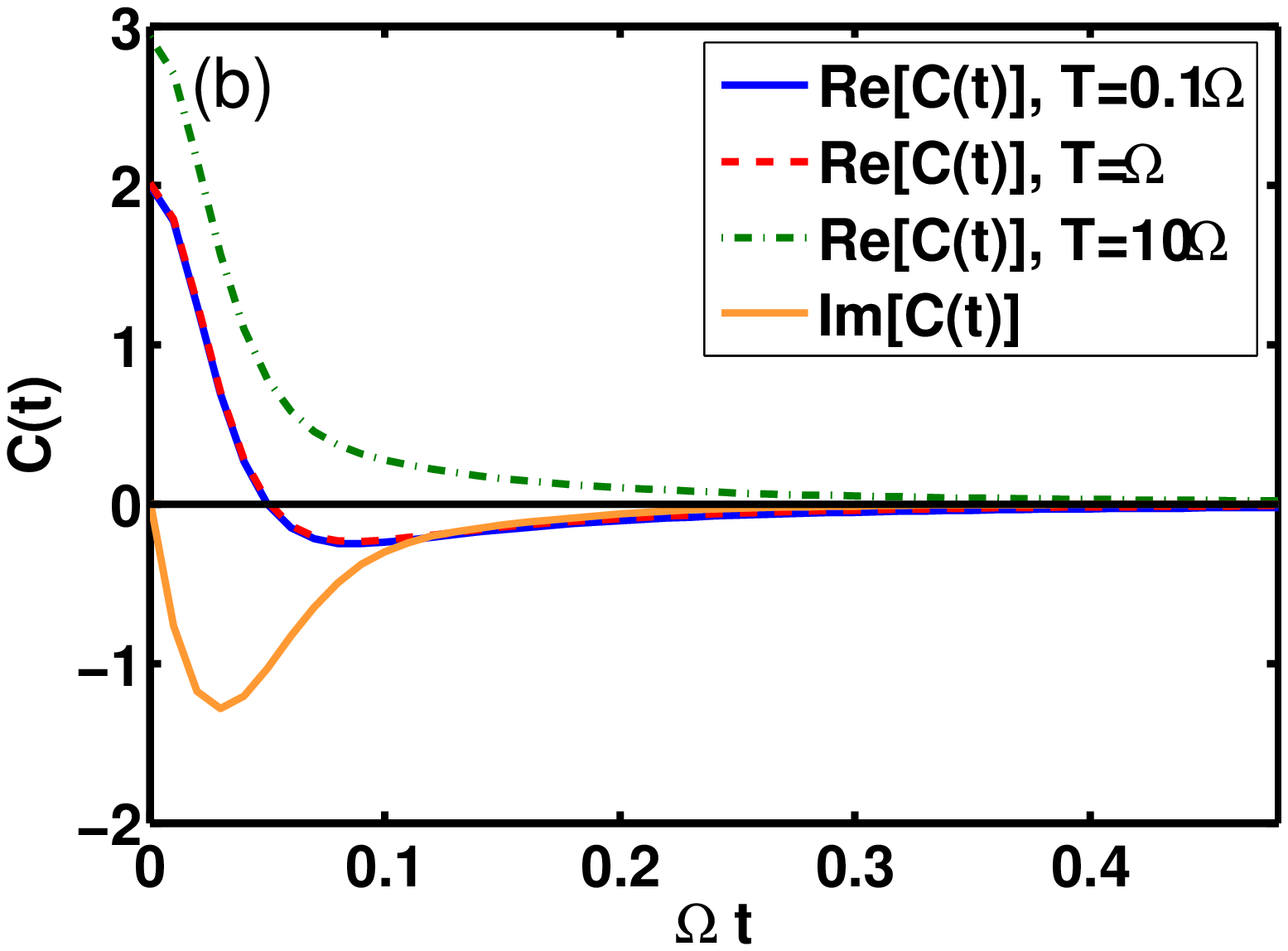}}}}
\caption{(color online) (a) Z-gate error versus time $t_f$ for different
  values of $\alpha$ and $T$ at $\omega_c=20\Omega$. 
The stopping criterion of the error
  threshold is set to $10^{-6}$ or when the number of iterations
  exceeds $3000$. (b) Corresponding bath CF's
  for $\alpha=0.01$. 
We set $k_B=1$ and $\hbar=1$ in all the figures
presented here and below.}
\label{error_vs_time_wc_20}
\end{figure}

\begin{figure}[tbp]
\mbox{\subfigure{\includegraphics[width=0.45\linewidth]{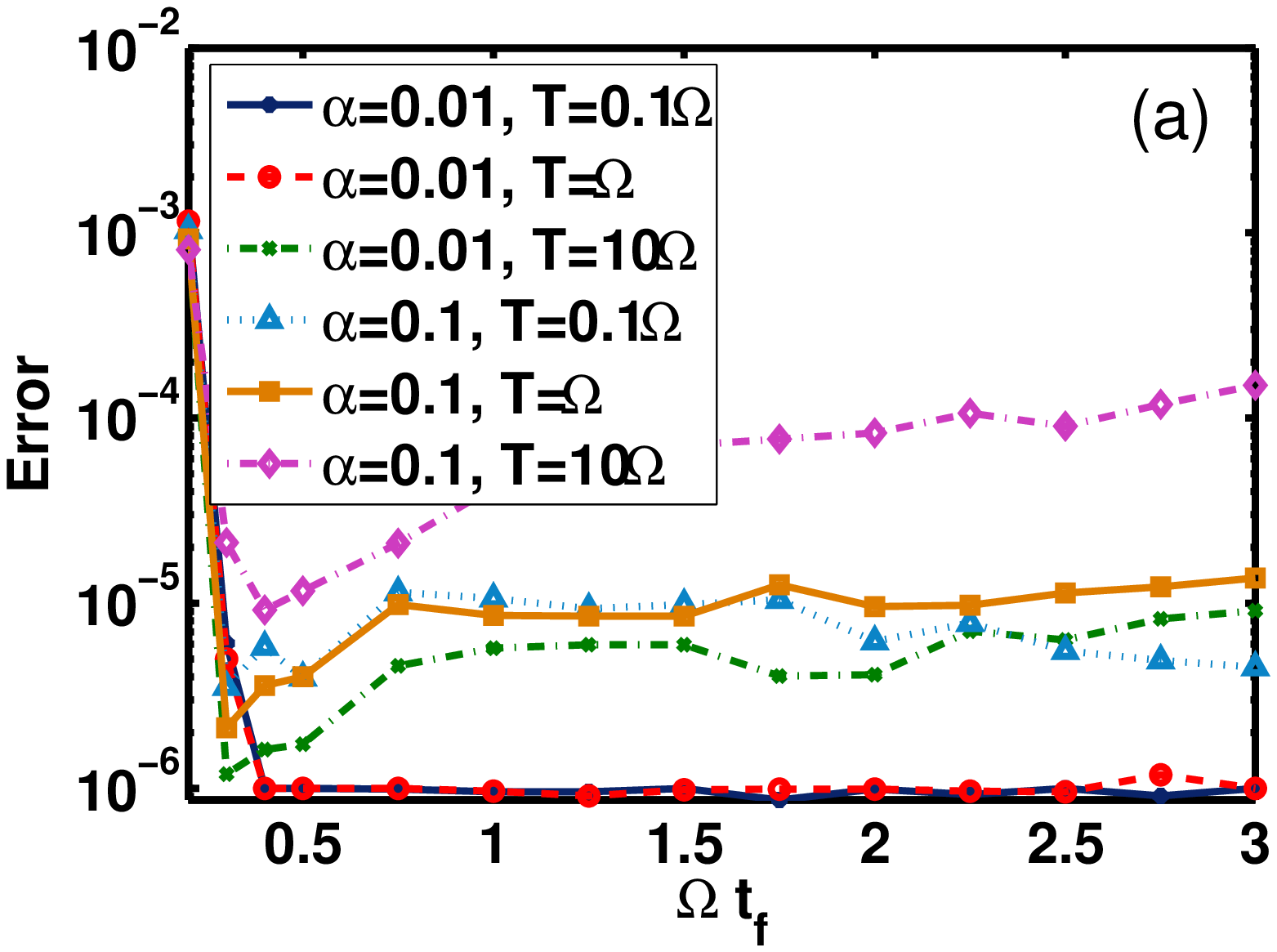}\subfigure{\includegraphics[width=0.45\linewidth]{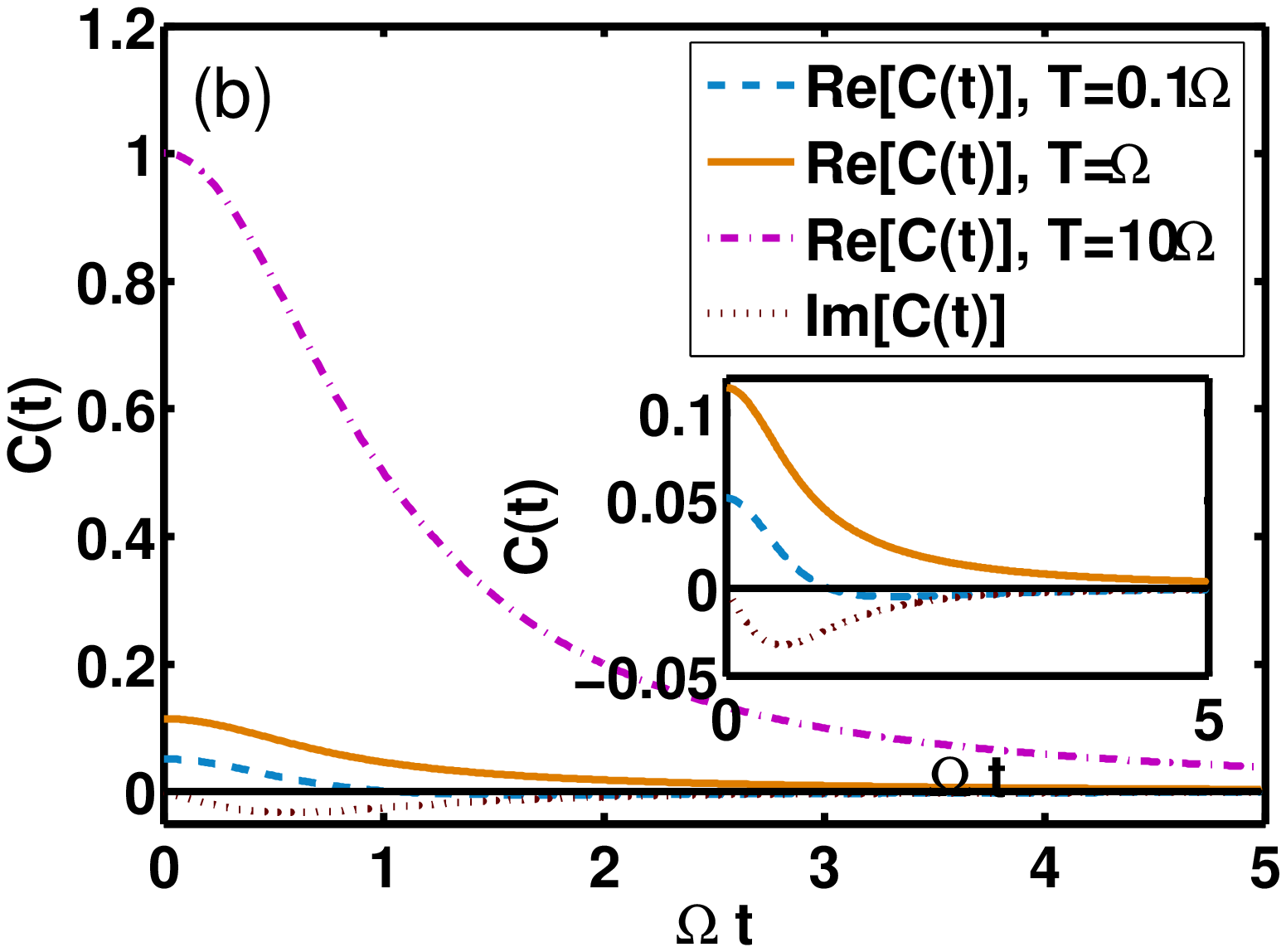}}}}
\caption{(Color online) (a) $Z$-gate error versus time $t_f$ for the same bath
  parameters and stopping criterion
  as in Fig.~\ref{error_vs_time_wc_20} except $\omega_c=\Omega$.
(b) Corresponding bath CF's  for $\alpha=0.1$. The lower three curves
are plotted in a smaller vertical-axis range in the inset.}
\label{error_vs_time_wc_1}
\end{figure}


\begin{figure}[tbp]
\mbox{\subfigure{\includegraphics[width=0.45\linewidth]{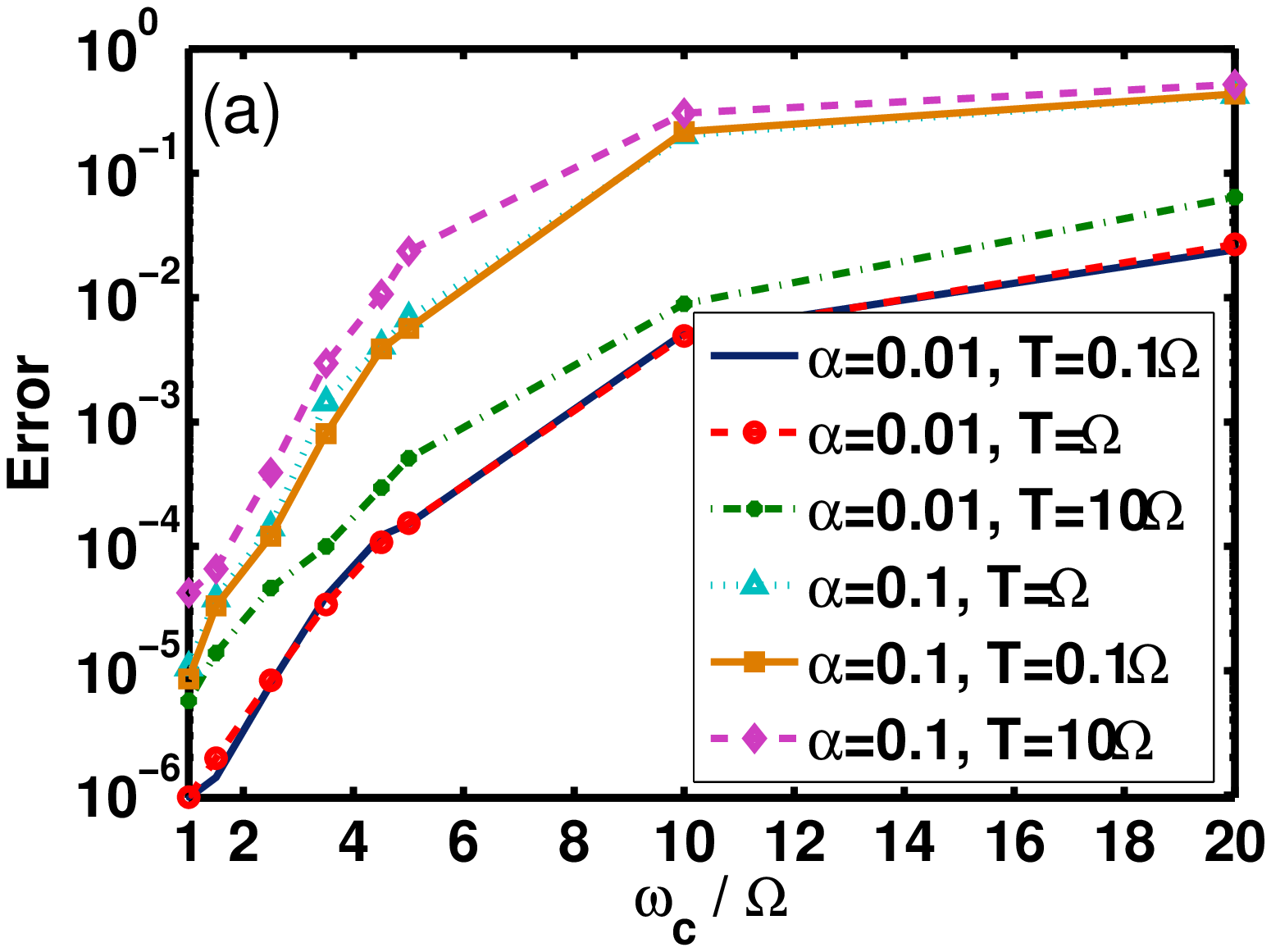}}\subfigure{\includegraphics[width=0.45\linewidth]{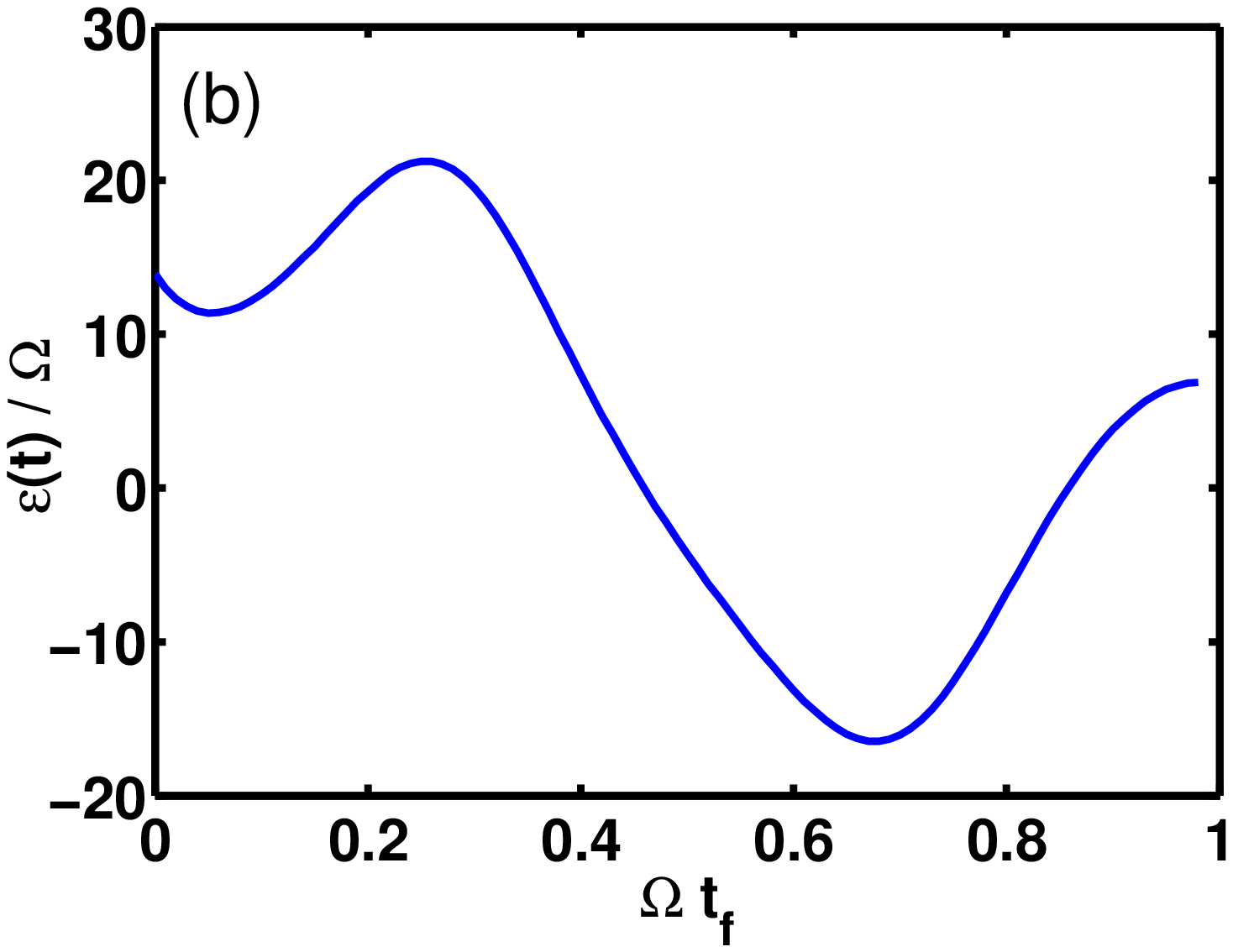}}}
\caption{(Color online)  
(a) $Z$-gate error versus cutoff $\omega_c$ for different
  values of $\alpha$ and $T$ with $t_f=\Omega^{-1}$.
(b) Optimal $Z$-gate control sequence for $\alpha=0.01$, $T=\Omega$,
$t_f=\Omega^{-1}$ and $\omega_c=\Omega$.} 
\label{error_vs_wc_tf_1}
\end{figure}


The ideal $Z$-gate performance as a function of the
gate operation time $t_f$  in the absence of the bath
is given in Fig.~\ref{ideal gate} 
with the stopping criterion of error set to $10^{-8}$. 
For all the calculations performed in this paper, 
we restrict the maximum control parameter $\varepsilon(t)\leq 30
\Omega$.
Excellent $Z$-gate performance can be achieved for pulse time $t_f\geq
0.3/\Omega$.
The corresponding optimal pulse sequence is shown in
the inset of Fig.~\ref{ideal gate}(a).   
In fact, 
a unitary $Z$-gate for pulse time 
$t_f > 0.3/\Omega$ with error limited only 
by machine precision can be achieved. 
Furthermore, if one does not impose any restriction on the maximum control
parameter strength,
an optimal perfect $Z$-gate
can be achieved for any finite period of time $t_f$. 
This is in contrast to the control pulse strategy for the
implementation of a (unitary) $Z$-gate in
\cite{Rebentrost2009}, where the gate operation time $t_f \geq
\pi/\Omega$, corresponding to the static
$\Omega \sigma_x$ inducing at least a full loop around the $x$ axis. 
One can see from the Bloch polarization vector 
evolution in Fig.~\ref{ideal gate}(b) that our optimal
control pulse strategy does not require a full loop around the $x$ axis
as compared with the state evolution in Fig.~2 (left panel) of
\cite{Rebentrost2009}, and thus can achieve a $Z$-gate in a
much shorter operation time 
limited primarily by the control parameter strength.

Figures \ref{error_vs_time_wc_20} and \ref{error_vs_time_wc_1} show
the errors of the $Z$-gate versus operation times  and the
corresponding bath CF's  
for different values of the bath
dimensionless damping constant $\alpha$ and temperature $T$
at cutoff frequencies
$\omega_c=20\Omega$ and $\omega_c=\Omega$, respectively.
Note that the imaginary part of the bath CF
does not depend on temperature $T$, so only  
one curve of the imaginary part is
plotted in Figs.~\ref{error_vs_time_wc_20}(b) 
and \ref{error_vs_time_wc_1}(b).
One can see that for $\omega_c=20\Omega$, 
the gate errors increase with the operation times $t_f$ for all the 
different bath parameter cases  
studied in Fig.~\ref{error_vs_time_wc_20}(a). 
The errors for the case of smaller $\alpha=0.01$ are about $10^{-3}$ at
the beginning and increase to $10^{-1}$ at longer operation times. 
As expected, the gate performance deteriorates as the value
of $\alpha$ increases.
The value of $\alpha$ plays a more important role 
than that of temperature in determining the amount of the error for the
$\omega_c=20\Omega$ case.
The bath CF at time $t=0$ for the high-temperature case of 
$k_BT=10\hbar\Omega$ 
is bigger than that for the low temperature case of $k_BT=\hbar\Omega$ by
a factor of only $1.5$ as shown in Fig.~\ref{error_vs_time_wc_20}(b).
However, 
the bath CF (effective decay rate) 
is directly proportional to $\alpha$.
Therefore, the gate errors
for the case $\omega_c=20\Omega$ shown in
Fig.~\ref{error_vs_time_wc_20}(a)
depend mainly on the value of $\alpha$. 
In all the previous 
investigations of quantum gate performance for both Markovian and
non-Markovian open quantum systems \cite{Rebentrost2009,potz2008,potz2009,Jirari2009}, 
the gate errors for a value
of $\alpha=0.01$ are about $10^{-3}$ or worse.
This is similar to what we found for the $\omega_c=20\Omega$ case. 
However, one of our main results, significantly 
different from previous investigations, is that  
for small $\omega_c$, it is possible to achieve high-fidelity $Z$-gates 
with errors smaller than $10^{-5}$ at large times, 
as shown in Fig.~\ref{error_vs_time_wc_1}(a).
This can be understood from the control-dissipation correlation, 
Eq.~(\ref{K_diss}), and the
memory effect of the non-Markovian environment.  
The bath CF's in Figs.~\ref{error_vs_time_wc_1}(a) and
\ref{error_vs_time_wc_20}(a) decay to zero on a
time scale of $t_B\approx \omega_c^{-1}$, called the bath correlation time. 
For $\omega_c=20\Omega$, the bath is very close to Markovian, 
and the dissipator (\ref{K_diss})
or decay rate approaches a constant value in a very
short time of $t_B\approx 0.05\Omega^{-1}$ and thus 
cannot be significantly suppressed by the external control sequence.   
On the other hand, for $\omega_c=\Omega$, the longer bath correlation
time of $t_B\approx \Omega^{-1}$ allows the optimal control sequence 
to have enough action time to counteract and suppress 
the contribution from the bath 
in the weak system-bath coupling case, and thus reduces the
gate error considerably. 
For $\alpha=0.01$ and temperature $k_BT\leq\hbar\Omega$, the gate error can be
kept about or smaller than $10^{-6}$. 
The temperature $T$ plays a similar role to the coupling strength when
$k_BT>\hbar\Omega$, as the amplitude ratio of the bath CF between the
$k_BT=10\hbar\Omega$ and $k_BT\leq\hbar\Omega$ cases is about
$10$. 
Even though the
gate errors for larger $\alpha$ and higher
$T$ increase with $t_f$,
the errors for all of the cases studied
except for the case of $\alpha=0.1$ and 
$T=10\Omega$ in Fig.~\ref{error_vs_time_wc_1}(a)
are smaller than $10^{-5}$, smaller than the error
threshold of $10^{-4}$ ($10^{-3}$) \cite{Preskill09}
required for fault-tolerant quantum
computation.  
We also perform calculations for an optimal identity gate 
and the behaviors of the gate errors versus operation times $t_f$ are
similar to that of the $Z$-gate in Fig.~\ref{error_vs_time_wc_1}(a).
Achieving a high-fidelity identity gate at long
times implies having the capability for arbitrary state preservation, 
i.e., storing arbitrary state robustly against the bath.  
The gate errors are expected to be much lower if one has independent control
over both the $\sigma_z$ and $\sigma_x$ terms in the qubit Hamiltonian
and if there is no restriction on the control parameter strengths.

Figure \ref{error_vs_wc_tf_1}(a) shows the Z-gate error versus bath
cutoff frequency 
$\omega_c$ at operation time $t_f=\Omega^{-1}$ for different values of
$\alpha$ and $T$. One can see that the gate error depends strongly on
the bath cutoff frequency. The error increases as $\omega_c$ becomes
bigger. For the weak coupling and low temperature cases
($\alpha=0.01$, $T\leq \Omega$), it is possible to reduce the error to
below $10^{-5}$ for $\omega_c\leq
2.5\Omega$. 
These indicate that the bath correlation time $t_B\approx
\omega_c^{-1}$ or memory effect plays
an important role in determining the gate error. 
Figure \ref{error_vs_wc_tf_1}(b) shows the optimal control
sequence
 for $\omega_c=\Omega$, $\alpha=0.01$, $k_BT=\hbar\Omega$ and
$t_f=\Omega^{-1}$.
The optimal control sequence that suppresses the decoherence induced
by the bath  
is totally different from that of the ideal unitary case in the inset of
Fig.~\ref{ideal gate}(a).

\section{Conclusion}
To conclude, a universal QOCT based on the Krotov method for a
time-nonlocal non-Markovian open quantum system has been
introduced and applied to obtain control sequences and gate errors for
$Z$ and identity gates.
Our study has yielded several computational and conceptual innovations:
(a) The QOCT that combines the Krotov method and an extended 
Liouville space quantum dissipation formulation that transforms the
nonlocal-in-time master equation to a set of coupled linear
local-in-time equations of motion is introduced 
to deal with non-Markovian open quantum systems. 
(b) The direct numerical decomposition of the bath CF into
multi-exponential form has a great computational 
advantage over the commonly used spectral density parametrization approach
\cite{Meier99,Yan01,Kleinekathofer04,Kleinekathofer06}.  
(c) The constructed QOCT, which retains the merits of the Krotov method,
is extremely efficient in dealing with the time-nonlocal non-Markovian
equation of motion. Compared to the calculations performed on a 40-node
SUN Linux cluster via the gradient-based approach to tackle the
nonlocal kernel directly \cite{potz2008}, the calculations using our
approach for a
similar problem can 
be performed on a typical laptop PC with ease, thus opening the way for
investigating two-qubit and many-qubit problems in non-Markovian
environments. 
(d) Our study of optimal control 
reveals the strong
dependence of the gate 
errors on the bath correlation time and exploits this 
non-Markovian memory effect for
high-fidelity quantum gate implementation and arbitrary state
preservation in an open quantum system.  
The presented QOCT has been shown to be a powerful tool, capable of
facilitating implementations of various quantum information tasks
against decoherence. The required information is knowledge of the 
bath or noise spectral density 
which is experimentally accessible 
by, for example, dynamical 
  decoupling noise spectroscopy techniques
  \cite{Yuge11,Suter11,Bylander11,Almog11,Biercuk11}.
We note here that  
not only our proposed method of QOCT but also dynamical decoupling
and other strategies for fighting decoherence require knowledge of the
spectral distribution of the noise (bath spectral density) in order to
improve the strategies and design effective and/or optimized control
sequences \cite{Suter07,Uhrig08,Biercuk09,Uys09,Gordon08,Clausen10}.
Thus using
the dynamical decoupling noise spectroscopy techniques
\cite{Yuge11,Suter11,Bylander11,Almog11,Biercuk11} to determine
the bath spectral densities experienced by the qubits and then
applying QOCT to find explicitly control sequences for quantum gate
operations for the non-Markovian open qubit system will yield fast
quantum gates with 
low invested energy and high fidelity .
By virtue of its generality and efficiency, our Krotov based QOCT method 
will find useful applications in many different branches of the sciences.  
Recent experiments on engineering external environments
\cite{Wineland00}, simulating open quantum systems \cite{Blatt11},
and observing non-Markovian dynamics \cite{Madsen11,Liu} could
facilitate the experimental realization of the QOCT in
non-Markovian open quantum systems in the near future.

\begin{acknowledgments}
HSG acknowledges support from the
National Science Council in Taiwan under Grant
No. 100-2112-M-002-003-MY3, 
from the National Taiwan University under Grants
No. 10R80911 and No. 10R80911-2, and
from the
focus group program of the National Center for Theoretical
Sciences, Taiwan.
\end{acknowledgments}

\appendix*
\section{Derivation of the time-nonlocal master equation}
\label{appendix}
Here we provide the derivation of the time-nonlocal non-Markovian
master equation (\ref{EOM0}) and (\ref{K_diss}) in the text. 
Following the assumption of factorized initial system-bath state $\rho_T(0)=\tilde\rho_T(0)=\rho(0)\otimes R_0$, the standard perturbative time-convolution (under only the Born approximation) master equation in the interaction picture takes the form of 
\begin{eqnarray}
\dot{\tilde{\rho}}(t)&=&
-\frac{i}{\hbar}{\rm Tr}_B[\tilde{H}_I(t),\rho(0)\otimes R_0]
\nonumber\\
&&-\frac{1}{\hbar^2}{\rm Tr}_B\int_0^t dt'[\tilde{H}_I(t),[\tilde{H}_I(t'),\tilde{\rho}(t')\otimes R_0]].
\label{interEoM}
\end{eqnarray}
Here $\tilde{\rho}(t)$ is the system density matrix
in the interaction picture,
$R_0=\exp(-H_B/k_B T)/{\rm Tr}_B[\exp(-H_B/k_B T)]$ is the initial
thermal reservoir density operator at temperature $T$, and
the system-bath interaction Hamiltonian in the interaction picture
in our spin-boson model can be written as
\begin{equation}
\tilde{H}_{I}(t)
=\tilde{\sigma}_x(t)B(t),
\label{HI}
\end{equation}
 where $\tilde{\sigma}_x(t)=U_S^\dagger(t)\sigma_x U_S(t)$  with
 $U_S(t)=T_+e^{\frac{i}{\hbar}\int_0^t H_S(t')dt'}$  being the system
 evolution operator and $T_+$ being the time-ordering operator, and
 $B(t)=\sum_q c_qb_qe^{-i\omega_q t}+c_qb_q^\dagger e^{i\omega_q
   t}$. 
Substituting  Eq.~(\ref{HI}) into Eq.~(\ref{interEoM}) and then
 tracing over the reservoir (bath or environment) degrees of freedom, one obtains 
\begin{eqnarray}
\dot{\tilde{\rho}}(t)&=&-\frac{1}{\hbar^2}\int^t_0{\rm Tr}_B[\tilde{\sigma}_x(t)B(t),[\tilde{\sigma}_x(t')B(t'),\tilde{\rho}(t')\otimes{R_0}]]dt'\nonumber\\
&=&-\frac{1}{\hbar^2}\int_0^t
dt'\left\{[\tilde{\sigma}_x(t)\tilde{\sigma}_x(t')\tilde{\rho}(t')
\right.\nonumber\\
&&\qquad\qquad\qquad\left. -\tilde{\sigma}_x(t')\tilde{\rho}(t')\tilde{\sigma}_x(t)]C(t-t')
\right.\nonumber\\
&&\qquad\qquad\quad\left.+[\tilde{\rho}(t')\tilde{\sigma}_x(t')\tilde{\sigma}_x(t)
\right. \nonumber\\
&&\qquad\qquad\qquad\left. -\tilde{\sigma}_x(t)\tilde{\rho}(t')\tilde{\sigma}_x(t')]C(t'-t)
\right\},
\end{eqnarray}
where the relation ${\rm Tr}_B[\tilde{H}_I(t)R_0]=0$ has been used to
eliminate the first-order term in Eq.~(\ref{interEoM}). 
The bath correlation function is defined as 
\begin{eqnarray}
C(t-t')
& \equiv & {\rm Tr}_B[B(t)B(t')R_0]\nonumber\\
&= & \int_0^\infty d\omega
J(\omega)\left\{[n(\omega)+1]e^{-i\omega(t-t')}
\right.\nonumber\\
&&\qquad \qquad \quad \left. +n(\omega) e^{i\omega (t-t')}\right\}
\nonumber\\
&=&\int_0^\infty d\omega J(\omega)\cos[\omega
(t-t')]\coth\left[\frac{\hbar\omega}{2k_BT}\right] 
\nonumber\\
&& -i\int_0^\infty d\omega
J(\omega)\sin[\omega (t-t')],
\label{bathCF}
\end{eqnarray}
where we have used the definition of $B(t)$ below Eq.~(\ref{HI}) and 
the definition of the spectral density
$J(\omega)=\sum_q|c_q|^2\delta(\omega-\omega_q)$ to obtain 
Eq.(\ref{bathCF}),
 and $n(\omega)$ is the canonical ensemble average occupation
number of the bath. 
Rotating back to the Schr\"{o}dinger picture, 
the nonlocal-in-time non-Markovian master equation for the system density matrix $\rho(t)$ takes the form
\begin{equation}
\dot{\rho}(t)=-\frac{i}{\hbar}[H_S(t),\rho(t)]
-\frac{i}{\hbar}\{[\sigma_x,\mathcal{K}(t)]
-[\mathcal{K}^\dagger(t),\sigma_x]\},    
\label{nonMar-rho-eq}
\end{equation} 
where 
\begin{equation}
\mathcal{K}(t)=-\frac{i}{\hbar}U_S(t)\left[ \int_0^t dt'
  C(t-t')\tilde{\sigma}_x(t') \tilde{\rho}(t')\right] U_S^\dagger(t),
\label{Kt1}
\end{equation}
and $\tilde{\rho}(t)=U_S^\dagger(t)\rho(t)U_S(t)$.
We can further rewrite Eqs.~(\ref{nonMar-rho-eq}) and (\ref{Kt1}) in a superoperator form as
\begin{equation}
\dot \rho(t)=\mathcal{L}_S(t)\rho(t)+\{\mathcal{L}_x\mathcal{K}(t)+[\mathcal{L}_x \mathcal{K}(t)]^\dagger\},
\label{EoM-NonM}
\end{equation}
and
\begin{eqnarray}
\mathcal{K}(t)=-\frac{i}{\hbar}\int_0^t dt' C(t-t')
\mathcal{U}_S(t,t')\sigma_x \rho(t').
\label{Kt}
\end{eqnarray}
Here the superoperators $\mathcal{L}_S(t)$ and $\mathcal{L}_x$ are
defined via their actions on an arbitrary operator $A$, respectively, as
$\mathcal{L}_S(t) A= -(i/\hbar)[H_S(t),A]$ and $\mathcal{L}_x A=
-(i/\hbar)[\sigma_x,A]$, and the system propagator superoperator
$\mathcal{U}_S(t,t')=T_+\exp\{\int_{t'}^td\tau\mathcal{L}_S(\tau)\}$.
Equations (\ref{EoM-NonM}) and (\ref{Kt}) are just 
Eqs.~(\ref{EOM0}) and (\ref{K_diss}) in
the text.


\begin{references}
\bibitem{Rabitz88}A. P. Peirce et al.,
  Phys. Rev. A 37, 4950 (1988).
\bibitem{Tannor92} D. J. Tannor et al., in {\it
    NATO ASIB Proc. 299: Time-Dependent Quantum Molecular Dynamics},
  edited by J. Broeckhove and L. Lathouwers (1992) pp. 347-360.

\bibitem{Kosloff02} J. P. Palao and R. Kosloff, Phys. Rev. Lett.,
  \textbf{89}, 188301 (2002); Phys. Rev. A,
  \textbf{68}, 062308 (2003).
\bibitem{Montangero2007} S. Montangero et al., 
  Phys. Rev. Lett., \textbf{99}, 170501 (2007).
\bibitem{Nielsen08} I. I. Maximov et al., 
  J. Chem. Phys. \textbf{128}, 184505 (2008).
\bibitem{Khaneja05} 
A. Sp\"orl et al.,
  Phys. Rev. A, \textbf{75}, 012302 (2007);
D.-B. Tsai, P.-W. Chen, and H.-S. Goan, Phys. Rev. A
  \textbf{79}, 060306 (R) (2009). 
\bibitem{Tannor11}R. Eitan et al., Phys. Rev. A,
  \textbf{83}, 053426 (2011).


\bibitem{Rebentrost2009} P. Rebentrost et al., Phys. Rev. Lett. \textbf{102}, 090401 (2009).
\bibitem{potz2008} M. Wenin and W. P\"otz, Phys. Rev. A \textbf{78},
  012358 (2008);
Phys. Rev. B \textbf{78},
  165118 (2008).  M. Wenin et al., J.
  Appl. Phys. \textbf{105}, 084504 (2009).
\bibitem{potz2009} R. Roloff and W. P\"otz, Phys. Rev. B \textbf{79},
  224516 (2009); M. Wenin and W. P\"otz, Phys. Rev. A \textbf{74},
  022319 (2006).
\bibitem{Jirari2009} H. Jirari, EuroPhys. Lett.
  \textbf{87}, 40003 (2009).
\bibitem{Gordon08} G. Gordon et al., Phys. Rev. Lett. \textbf{101},
  010403 (2008).
\bibitem{Clausen10}J. Clausen et al., 
  Phys. Rev. Lett. \textbf{104}, 040401 (2010).
\bibitem{West10}J. R. West et al.,
  Phys. Rev. Lett. \textbf{105}, 230503 (2010) and references therein.
\bibitem{krotov1996} V. F. Krotov, in {\it Global methods in optimal
    control theory}, edited by Z. N. Earl and J. Taft (Marcel Dekker,
  New York, 1996).



\bibitem{Schmidt11}R. Schmidt  et al., Phys. Rev. Lett. \textbf{107},
  130404 (2011). 


\bibitem{Breuer02} H.P.~Breuer and F.~Petruccione, {\it The Theory of
    Open Quantum Systems} (Oxford University Press, Oxford, 2002).

\bibitem{Hu92}B. L. Hu, J. P. Paz, and Y. Zhang, Phys. Rev. D 45, 2843
  (1992).

\bibitem{Diosi98} L. Di\'{o}si, N. Gisin, and W. T. Strunz,
  Phys. Rev. A \textbf{58}, 1699 (1998).

\bibitem{Strunz04}W. T. Strunz and T. Yu, Phys. Rev. A. \textbf{69}, 052115 (2004).

\bibitem{Yan09} R. X. Xu et al., J. Chem. Phys. \textbf{130}, 074107
  (2009).
\bibitem{Chang10} K. W. Chang and C. K. Law
Phys. Rev. A \textbf{81}, 052105 (2010).
\bibitem{Yu10} J. Jing and T. Yu
Phys. Rev. Lett. \textbf{105}, 240403 (2010).





\bibitem{Meier99} C. Meier et al.,
  J. Chem. Phys. \textbf{111}, 3365 (1999).
\bibitem{Yan01} F. Shuang et al., J. Chem. Phys. \textbf{114},
  3868 (2001); R. X. Xu and Y. J. Yan, {\it ibid.} \textbf{116}, 9196 (2002). 
\bibitem{Kleinekathofer04} U. Kleinekath\"ofer,
J. Chem. Phys. \textbf{121}, 2505 (2004).
\bibitem{Kleinekathofer06} S. Welack et al., 
 J. Chem. Phys. \textbf{124}, 044712 (2006).



\bibitem{ER}F. Poyatos et al., Phys. Rev. Lett. {\bf 77}, 4728 (1996);
  C. J. Myatt et al., Nature (London) {\bf 403}, 269 (2000);
  F. O. Prado et al., Phys. Rev. Lett. {\bf 102}, 073008 (2009).

\bibitem{ICE} A. Pechen et al., Phys. Rev. A {\bf 73}, 062102 (2006);
  R. Romano et al., Phys. Rev. A {\bf 73}, 022323 (2006);
  Phys. Rev. Lett. \textbf{97}, 080402 (2006). 


\bibitem{Preskill09} P. Aliferis et al., 
Phys. Rev. A \textbf{79}, 012332 (2009). 

\bibitem{Yuge11} T. Yuge, S. Sasaki, and Y. Hirayama,
  Phys. Rev. Lett. \textbf{107}, 170504 (2011).
\bibitem{Suter11} G. A. \'Alvarez and D. Suter, Phys. Rev. Lett. \textbf{107}, 230501
    (2011). 
\bibitem{Bylander11} J. Bylander et al., Nat. Phys. \textbf{7}, 565 (2011).
\bibitem{Almog11} I. Almog, Y. Sagi, G. Gordon, G. Bensky, G. Kurizki, and
N. Davidson, J. Phys. B: At. Mol. Opt. Phys. \textbf{44}, 154006 (2011).
\bibitem{Biercuk11} M. J. Biercuk,  Nat. Phys. \textbf{7}, 525 (2011).

\bibitem{Suter07} J. Zhang, X. Peng, N. Rajendran, and D. Suter, Phys. Rev.
A 75, 042314 (2007).
\bibitem{Uhrig08} G. S. Uhrig, New J. Phys. 10, 083024 (2008).
\bibitem{Biercuk09} M. J. Biercuk et al., Nature (London) 458, 996 (2009).
\bibitem{Uys09} H. Uys, M. J. Biercuk, and J. J. Bollinger, Phys. Rev. Lett.
103, 040501 (2009).

\bibitem{Wineland00} C. J. Myatt et al., Nature (London) \textbf{403}, 269 (2000).
\bibitem{Blatt11} J. T. Barreiro et al., Nature (London) \textbf{470}, 486 (2011).
\bibitem{Madsen11}K. H. Madsen et al.,
  Phys. Rev. Lett. \textbf{106}, 233601 (2011).
\bibitem{Liu} B. H. Liu et al., Nat. Phys. \textbf{7}, 931 (2011).

\end{references}
\end{document}